\documentclass[pra,aps,superscriptaddress,twocolumn,amssymb,amsmath,longbibliography]{revtex4-2}
\usepackage{graphicx}
\usepackage{dcolumn}
\usepackage{bm}
\usepackage[mathlines]{lineno}
\usepackage[utf8]{inputenc}
\usepackage[T1]{fontenc}
\usepackage{mathptmx}
\usepackage{float}

\begin{document}

\title{A magnetic levitation based low-gravity simulator with an unprecedented large functional volume\vspace{7pt}}

\author{Hamid Sanavandi}
\author{Wei Guo}
\email[Corresponding author: ]{wguo@magnet.fsu.edu}
\affiliation{National High Magnetic Field Laboratory, 1800 East Paul Dirac Drive, Tallahassee, Florida 32310, USA}
\affiliation{Mechanical Engineering Department, FAMU-FSU College of Engineering, Florida State University, Tallahassee, Florida 32310, USA}
\date{\today}

\begin{abstract}
Low gravity environment can have a profound impact on the behaviors of biological systems, the dynamics and heat transfer of fluids, and the growth and self-organization of materials. Systematic research on the effects of gravity is crucial for advancing our knowledge and for the success of space missions. Due to the high cost and the limitations in the payload size and mass in typical spaceflight missions, ground-based low-gravity simulators have become indispensable for preparing spaceflight experiments and for serving as stand-alone research platforms. Among various simulator systems, the magnetic levitation based simulator (MLS) has received long-lasting interests due to its easily adjustable gravity and practically unlimited operation time. However, a recognized issue with MLSs is their highly non-uniform force field. For a solenoid MLS, the functional volume $V_{1\%}$, where the net force results in an acceleration less than 1\% of the Earth's gravity $g$, is typically a few microliters ($\mu L$) or less. In this work, we report an innovative MLS design that integrates a superconducting magnet with a gradient-field Maxwell coil. Through an optimization analysis, we show that an unprecedented $V_{1\%}$ of over 4,000 $\mu L$ can be achieved in a compact coil with a diameter of 8 cm. We also discuss how such an MLS can be made using existing high-$T_c$ superconducting materials. When the current in this MLS is reduced to emulate the gravity on Mars ($g_M=0.38g$), a functional volume where the gravity varies within a few percent of $g_M$ can exceed 20,000 $\mu L$. Our design may break new ground for various exciting future low-gravity research.

\end{abstract}
\maketitle

\section*{Introduction}\label{SecI}
\noindent Reduced gravity is known to have important effects on various biological and physical systems. For instance, a weightless environment may prohibit cell culture growth~\cite{Hammond-1999-NM} and may cause cellular stressor and bone loss that can negatively impact astronauts' health~\cite{White-2001-Nature,Williams-2009-CMAJ,Stavnichuk-2020-npj}. In fluid systems, reduced gravity can significantly affect the sloshing dynamics of cryogenic propellants in spacecrafts~\cite{Snyder-2001-Cryogenics}, the surface oscillation of liquid drops~\cite{Nelson-1972-JGR}, bubble cavitation~\cite{Obreschkow-2006-PRL} and boiling heat transfer in fluids~\cite{Siegel-1967-book,Ohta-2013-EHT}. In material science, the potential of reduced gravity in growing tissues~\cite{Unsworth-1998-NM} and crystals~\cite{Strelov-2014-CP} and for materials processing~\cite{Matisak-1997-JCG} has been recognized. Conducting systematic research to understand the mechanism of gravity in these diverse systems will undoubtedly advance our knowledge. Furthermore, various programs initiated recently by public space agencies and private organizations~\cite{NAP-2006, NASA-2015, Musk-2018-NS} aiming at long-term human habitation of the Moon and Mars have further strengthened the needs of experimental research in low gravity environment.

The ideal microgravity condition can be achieved in spaceflight experiments conducted during space-shuttle missions~\cite{Vandenbrink-2016-PlantC} and at space stations~\cite{Evans-2009-NASATP}. However, these experiments are limited by the high cost and the small payload size and mass~\cite{Ferranti-2021-AppSci}. The fact that the astronauts have to conduct the experiments instead of the trained scientists also put constraints on the design of the experiments. For these reasons, researchers have devoted great efforts in developing ground-based low-gravity simulators. One major category, which utilizes free fall to generate near-zero gravity, includes drop towers~\cite{Selig-2010-MST,Liu-2016-SR}, parabolic aircrafts~\cite{Melnikov-2008-MST,Shelhamer-2016-JAP}, sounding rockets~\cite{Fuhrmann-2008-MST}, and suborbital rocketry~\cite{Ohta-1997-ASME}. Despite their usefulness, a known limitation of these facilities is the relatively short low-gravity duration (i.e., from several seconds to a few minutes~\cite{Sundaresan-2006-JCMM}), which makes them unsuitable for experiments that require long observation times~\cite{Pletser-2004-ActaA}. In biological and medical research, rotational facilities such as clinostat machines~\cite{Dedolph-1971-PlantPhy,Herranz-2013-Astrobiology}, rotating wall vessels~\cite{Carlsson-2003-BBAMCR}, and random positioning machines~\cite{Wuest-2015-BiomedRes} are also adopted to achieve a small time-averaged gravity vector~\cite{Hoson-1997-Planta,Brown-1976-PlantPhy}. Although these simulators are convenient, they do not produce a genuine low-gravity environment and can generate unwanted centrifugal forces and circulating flows in the samples~\cite{Hoson-1997-Planta,Brown-1976-PlantPhy,Albrecht-2007-ASGSR}.

On the other hand, magnetic field-gradient levitation of various diamagnetic materials has been demonstrated~\cite{Hill-2008-PRL,Brooks-2000-JAP,Weilert-1996-PRL}. Even living organisms have been successfully levitated~\cite{Berry-1997-EJP,Valles-1997-Biophys,Liu-2010-ASR,Geim-1998-PhysToday,Schenck-2000-JMRI}, and there is no evidence of any cumulative harmful effects due to the field exposure~\cite{Geim-1998-PhysToday,Schenck-2000-JMRI,High-2000-JMRI}. Compared to other low-gravity simulator systems, a magnetic levitation based simulator (MLS) offers unique advantages, including low cost, easy accessibility, adjustable gravity, and practically unlimited operation time~\cite{Berry-1997-EJP,Valles-1997-Biophys,Valles-2005-ASR}. However, a known issue with MLSs is their highly non-uniform force field around the levitation point. If we define a 0.01-$g$ functional volume $V_{1\%}$ where the net force results in an acceleration less than 1\% of the Earth's gravity $g$, $V_{1\%}$ is typically less than a few microlitres ($\mu L$) for conventional solenoid MLSs. Although diamagnetic samples with sizes larger than $V_{1\%}$ can be levitated, a stress field caused by the residue force inside the samples can compromise the measurement results. Despite some limited efforts in designing MLSs for improved functional volumes~\cite{Chow-2011-IEEE, Kustler-2012-IEEE, Kuznetsov-2017-JMMM}, a major progress is still lacking. Furthermore, the high energy consumption rate of conventional resistive solenoid MLSs is also concerning. For instance, 4 MW electric power is required to levitate a frog using a resistive solenoid MLS~\cite{Berry-1997-EJP}.

In this paper, we report an innovative MLS design which consists of a gradient-field Maxwell coil placed in the bore of a superconducting (SC) magnet. By optimizing the SC magnet's field strength and the current in the Maxwell coil, we show that an unprecedented $V_{1\%}$ of over 4,000 $\mu L$ can be achieved in a compact coil of 8 cm in diameter. This optimum $V_{1\%}$ increases with the size and the field strength of the MLS. We then discuss how such a MLS can be made using existing high-$T_c$ superconducting materials so that long-time operation with minimal energy consumption can be achieved. To further demonstrate the usefulness of this MLS, we also consider reducing its current and the field strength to emulate the gravity on Mars ($g_M=0.38g$). It turns out that a functional volume over 20,000 $\mu L$ can be produced, in which the gravity only varies within a few percent of $g_M$. Our design concept may break new ground for exciting applications of MLSs in future low-gravity research.

\section*{Results}\label{SecII}
\noindent To aid the discussion of our MLS design, we first introduce the fundamentals of magnetic levitation using a solenoid magnet. Following this discussion, we will present the details of our innovative MLS design concept.

\subsection*{Levitation by a solenoid magnet}\label{SecIIa}
\noindent The concept of magnetic levitation can be understood by considering a small sample (volume $\Delta V$) placed in a static magnetic field $\mathbf{B}(\mathbf{r})$. Due to the magnetization of the sample material, the energy of the magnetic field increases by~\cite{Jackson-1999-book}
\begin{equation}
\Delta E_B=\frac {-\chi B^2(\mathbf{r})}{2\mu_0 (1+\chi )}\Delta V,
\end{equation}
where $\chi$ is the magnetic susceptibility of the sample material and $\mu_0$ is the vacuum permeability. For diamagnetic materials with negative $\chi$, $\Delta E_B$ is positive and therefore it requires energy to insert a diamagnetic sample into the $\mathbf{B}(\mathbf{r})$ field. Counting in the gravity effect, the total potential energy associated with the sample per unit volume can be written as:
\begin{equation}
E(\mathbf{r})=\frac {-\chi B^2(\mathbf{r})}{2\mu_0 (1+\chi )}+\rho gz,
\end{equation}
where $\rho$ is the material density and $z$ denotes the vertical coordinate. This energy leads to a force per unit volume acting on the sample as:
\begin{equation}
\mathbf{F}=-\mathbf{\nabla} E(\mathbf{r})=\frac {\chi}{\mu_0 (1+\chi )}\mathbf{B}\cdot\mathbf{\nabla}\mathbf{B}-\rho g \hat{e}_z.
\label{F-eqn}
\end{equation}
For an appropriate non-uniform magnetic field, the vertical component of the field-gradient force (i.e., the first term on the right side in Eq.~(\ref{F-eqn})) may balance the gravitational force at a particular location, i.e., the levitation point. Sample suspension can therefore be achieved at this point.

\begin{figure}[t]
\centering
\includegraphics[width=1\linewidth]{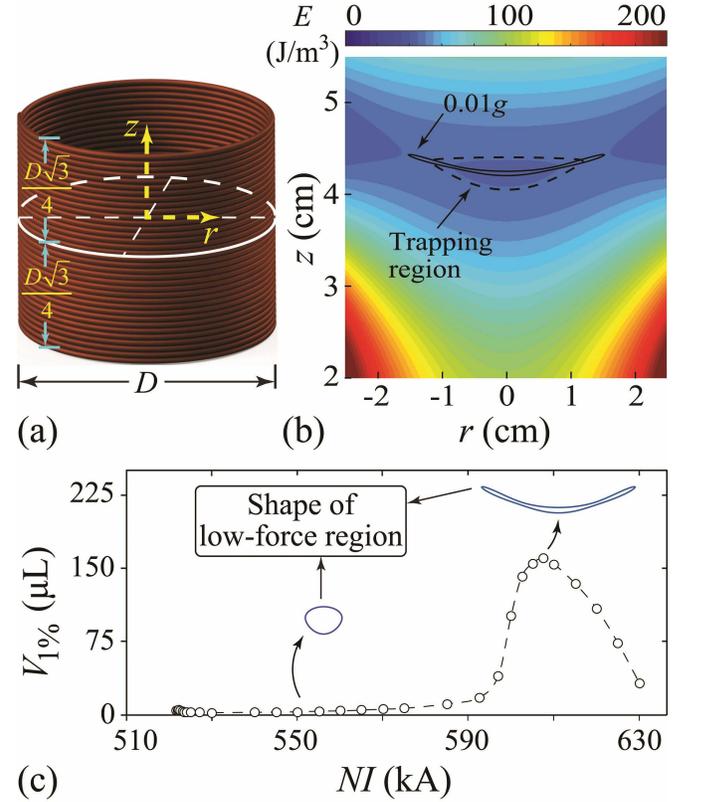}
\caption{(a) Schematic of a solenoid with a diameter of $D=8$ cm and a height of $\sqrt{3}D/2$. (b) Calculated specific potential energy $E(\mathbf{r})$ of a small water sample placed in the magnetic field. The turn-current $NI$ of the solenoid is 607.5 kA. The origin of the coordinates is at the center of the solenoid. The dashed contour denotes the boundary of the trapping region, and the solid contour shows the low-force region (i.e., acceleration less than $0.01g$). (c) The functional volume $V_{1\%}$ (i.e., overlapping volume of the two contours) versus the turn-current $NI$. Representative shapes of the low-force region are shown.}
\label{Fig1}
\end{figure}

In order to attain a stable levitation, the specific potential energy $E$ must have a local minimum at the levitation point so the sample cannot stray away. Since $E$ depends on the material properties besides the $\mathbf{B}(\mathbf{r})$ field, we need to specify the sample material. Considering the fact that water has been utilized in a wide range of low-gravity researches~\cite{Konishi-2015-IJHMT,Straub-2001-AHT,Tegart-1975-NASA} and is also the main constituent of living cells and organisms~\cite{Cooper-2007-ASMpress}, we adopt the water properties at ambient temperature~\cite{Eisenberg-2005-book} (i.e., $\chi=-9.1\times10^{-6}$ and $\rho=10^3$ kg/$m^3$) in all subsequent analyses. To see the effect of the $\mathbf{B}(\mathbf{r})$ field, we consider an solenoid with a diameter of $D=8$ cm and a height of $\sqrt{3}D/2$, as shown in Fig.~\ref{Fig1}~(a). These dimensions are chosen to match the size of the MLS that we will discuss in later sections. For a solenoid with $N$ turns and with an applied current $I$, $\mathbf{B}(\mathbf{r})$ can be calculated using a known integral formula that depends on the product $NI$ (see details in the Method section). $E(\mathbf{r})$ in the full space can then be determined.

In Fig.~\ref{Fig1}~(b), we show the calculated $E(\mathbf{r})$ near the top opening of the solenoid when a turn-current of $NI=607.5$ kA is applied. In general, $E$ is high near the solenoid wall due to the strong $\mathbf{B}$ field there. Slightly above the solenoid, there is a trapping region (enclosed by the dashed contour) in which $E$ decreases towards the region center. When a water sample is placed in this region, it moves towards the region center where the net force is zero, i.e., the levitation point. We have also calculated the specific force field using Eq.~(\ref{F-eqn}). The solid contour in Fig.~\ref{Fig1}~(b) denotes the low-force region in which the net force corresponds to an acceleration less than $0.01g$. The \emph{overlapping volume} of the trapping region and the low-force region is defined as our functional volume $V_{1\%}$ where the sample not only experiences weak residue forces but also remain trapped. In Fig.~\ref{Fig1}~(c), we show the calculated $V_{1\%}$ as a function of $NI$. The trapping region emerges only above a threshold turn-current of about $NI=520$ kA. As $NI$ increases, $V_{1\%}$ first remains small (i.e., a few $\mu L$) and has a shape like a inverted raindrop. When $NI$ is above about 600 kA, $V_{1\%}$ grows rapidly and peaks at $NI=607.5$ kA before it drops with further increasing $NI$. In the peak regime, $V_{1\%}$ has a highly anisotropic shape due to the non-uniform force field, which makes it unsuitable for practical applications despite the enhanced $V_{1\%}$ value. The required extremely large turn-current also presents a great challenge.

\begin{figure}[t]
\centering
\includegraphics[width=1\linewidth]{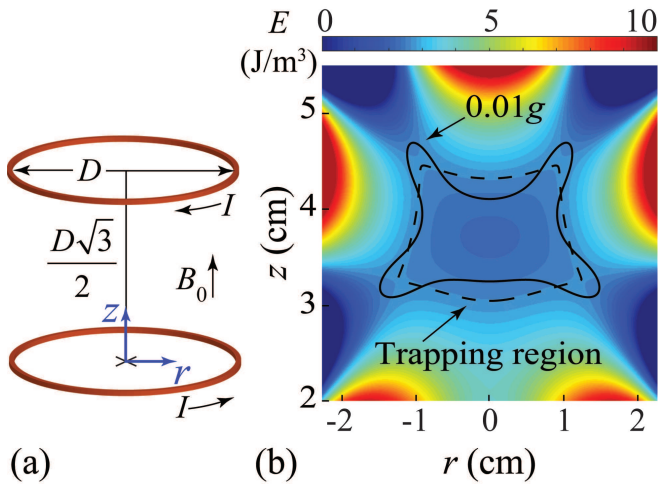}
\caption{(a) Schematic of the gradient-field Maxwell coil with a diameter $D=8$ cm in the presence of an applied uniform field $\mathbf{B}_0$. b) Calculated specific potential energy $E(\mathbf{r})$ of a small water sample placed in the magnetic field for $I=112.6$ kA and $B_0=24$ T. The origin of the coordinates is at the center of the bottom current loop. The black dashed contour denotes the boundary of the trapping region, and the black solid contour shows the low-force region (i.e., acceleration less than $0.01g$).}
\label{Fig2}
\end{figure}

\subsection*{Concept and performance of our MLS}\label{SecIIb}
\noindent To increase $V_{1\%}$, the key is to produce a more uniform field-gradient force to balance the gravitational force such that the net force remains low in a large volume. Base on Eq.~(\ref{F-eqn}), this can be achieved if we have a nearly uniform $\mathbf{B}$ field and at the meanwhile the field gradient is almost constant in the same volume. These two seemingly irreconcilable conditions can be satisfied approximately. The idea is to combine a strong uniform field $\mathbf{B}_0$ and a weak field $\mathbf{B}_1(\mathbf{r})$ that has a fairly constant field gradient $\mathbf{\nabla}\mathbf{B}_1$. In this way, the total field $\mathbf{B}=\mathbf{B}_0+\mathbf{B}_1\simeq\mathbf{B}_0$ is approximately uniform and its gradient $\mathbf{\nabla}\mathbf{B}\simeq\mathbf{\nabla}\mathbf{B}_1$ can also remain nearly constant.

\begin{figure}[t]
\centering
\includegraphics[width=1\linewidth]{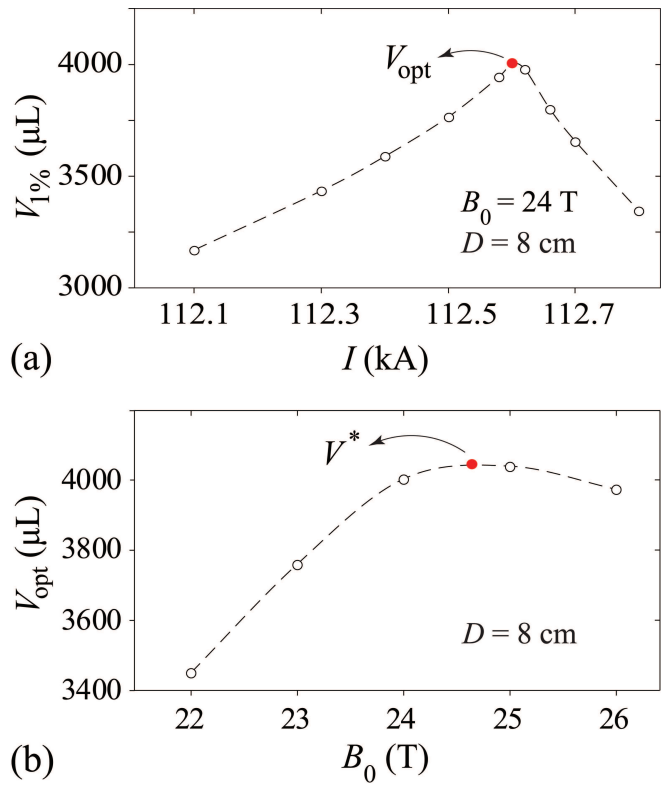}
\caption{(a) Calculated $V_{1\%}$ versus the loop current $I$ for the coil shown in Fig.~\ref{Fig2} with $B_0=24$ T. The largest $V_{1\%}$ is denoted as $V_{\mathrm{opt}}$. (b) The obtained $V_{\mathrm{opt}}$ as a function of $B_0$. The overall maximum $V_{\mathrm{opt}}$ is denoted as $V^*$, and the corresponding coil current and base field are designated as $I^*$ and $B^*_0$, respectively.}
\label{Fig3}
\end{figure}

The uniform field $\mathbf{B}_0$ can be produced in the bore of a superconducting solenoid magnet. Indeed, for superconducting magnets used in magnetic resonance imaging applications, spatial uniformity of the field better than a few parts per million (ppm) in a space large enough to hold a person has became standard~\cite{Sciver-2002-PT,Dixon-2005-IEEE}. The recently built 32-T all-superconducting magnet at the National High Magnetic Field Laboratory (NHMFL) further proves the feasibility of producing strong uniform fields using cutting-edge superconducting technology~\cite{Markiewicz-2011-IEEE}. As for the $\mathbf{B}_1$ field, we propose to produce it using a gradient-field Maxwell coil~\cite{Maxwell-1873-Book}. As shown in Fig.~\ref{Fig2}~(a), such a coil consists of two identical current loops (diameter $D$) placed coaxially at a separation distance of $\sqrt{3}D/2$. The current in the top loop is clockwise (viewed from the top) while the current in the bottom loop is counterclockwise. It was first demonstrated by Maxwell that such a coil configuration could produce a highly uniform field gradient in the region between the two loops~\cite{Maxwell-1873-Book}.

The $\mathbf{B}_1(\mathbf{r})$ generated by the gradient-field Maxwell coil can be calculated using the Biot-Savart law~\cite{Jackson-1999-book} (see details in the Method section), from which the specific potential energy $E$ for an inserted water sample can again be determined. As an example, we show in Fig.~\ref{Fig2}~(b) the calculated $E(\mathbf{r})$ profile for a coil with $D=8$ cm and with an applied current of $I=112.6$ kA in the presence of a uniform field $B_0=24$ T. Again, we use the dashed contour and the solid contour to show, respectively, the trapping region and the $0.01g$ low-force region. By evaluating the overlapping volume of the two regions, we obtain $V_{1\%}=4,004$ $\mu L$. More importantly, this functional volume is much more isotropic as compared to that in Fig.~\ref{Fig1}~(b), which makes it highly desirable in practical applications.

To optimize the coil current $I$ and the base field $B_0$, we have conduct further analyses. First, for a fixed $B_0$, we vary the coil current $I$. Representative results at $B_0=24$ T are shown in Fig.~\ref{Fig3}~(a). It is clear that $V_{1\%}$ peaks at about $I=112.6$. We denote this peak value as $V_{\mathrm{opt}}$. The decrease of $V_{1\%}$ at large $I$ is caused by the fact that the field $\mathbf{B}_1$ generated by the coil is no longer much smaller than the base field $\mathbf{B}_0$, which impairs the uniformity of the field-gradient force. Next, we vary the base field strength $B_0$ and determine the corresponding $V_{\mathrm{opt}}$ at each $B_0$. The result is shown in Fig.~\ref{Fig3}~(b). It turns out that there exists an optimum base field strength of about 24.7 T (denoted as $B^*_0$), where an overall maximum functional volume (denoted as $V^*$) of about 4,050 $\mu L$ can be achieved. This volume is comparable to those of the largest water drops adopted in the past spaceflight experiments~\cite{Fichtl-1987-Spacelab3,Tegart-1975-Skylab}.

\begin{figure}[htb]
\centering
\includegraphics[width=1\linewidth]{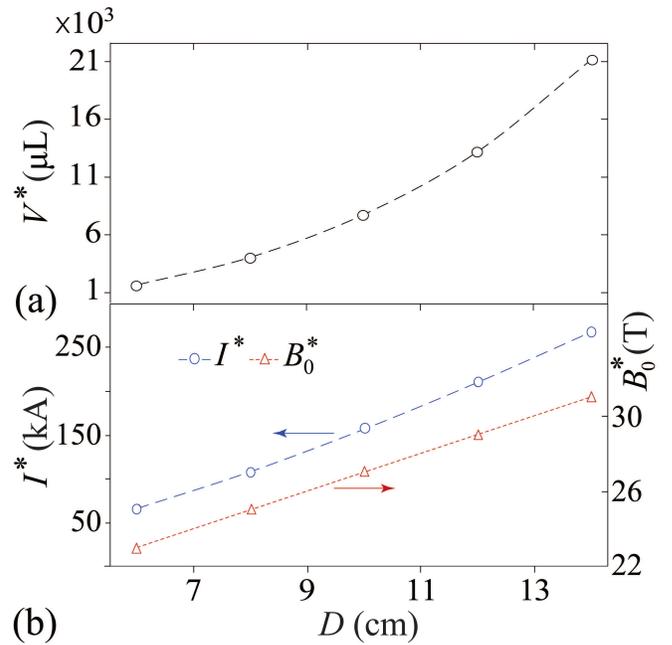}
\caption{(a) The maximum functional volume $V^*$ for coils with different diameters $D$. (b) The required optimum $I^*$ and $B^*_0$ to achieve $V^*$ versus the coil diameter $D$.}
\label{Fig4}
\end{figure}

The above analyses assumed a fixed coil diameter $D=8$ cm. When $D$ varies, the maximum functional volume $V^*$ and the corresponding MLS parameters (i.e., $I^*$ and $B^*_0$) should also change. To examine the coil size effect, we have repeated the aforementioned analyses with a number of coil diameters. The results are collected in Fig.~\ref{Fig4}. As $D$ increases from $6$ cm to $14$ cm, the maximum functional volume $V^*$ increases from about 1,500 $\mu L$ to over 21,000 $\mu L$, i.e., over 14 times. At the meanwhile, the required coil current $I^*$ and the base field strength $B^*_0$ increase almost linearly with $D$ by factors of about 4 and 1.3, respectively. This analysis suggests that it is advantageous to have a larger coil provided that the desired $I^*$ and $B^*_0$ can be achieved.

\section*{Discussion}\label{SecIII}
\subsection*{Practical design considerations}\label{SecIIIa}
\begin{figure*}[t]
\centering
\includegraphics[width=0.95\linewidth]{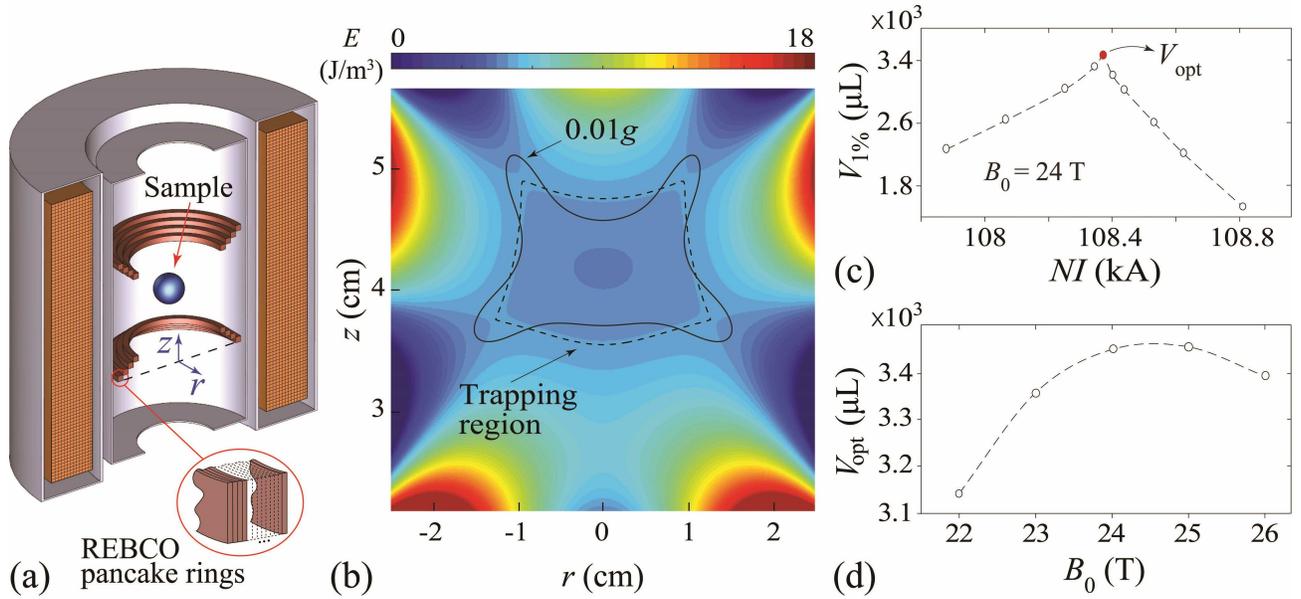}
\caption{(a) Schematic of a practical MLS setup that consists of a 24-T superconducting magnet with four sets of gradient-field Maxwell coils made of REBCO pancake rings. The averaged diameter of the pancake rings is about 8 cm. (b) Calculated specific potential energy $E(\mathbf{r})$ for a small water sample place in this MLS with a total turn-current $NI=108.37$ kA. The origin of the coordinates is at the center of the lowest pancake ring. The dashed contour denotes the boundary of the trapping region and the solid contour shows the $0.01g$ low-force region. (c) Calculated $V_{1\%}$ versus the turn-current $NI$ at $B_0=24$ T. The peak $V_{1\%}$ is denoted as $V_{\mathrm{opt}}$. (b) The obtained $V_{\mathrm{opt}}$ as a function of $B_0$.}
\label{Fig5}
\end{figure*}

\noindent The MLS concept that we have presented requires an applied current of the order $10^2$ kA in both loops of the gradient-field Maxwell coil. A natural question is whether this is practical. One may consider to make the loop using a thin copper wire with $10^3$ turns so that a current of the order $10^2$ A in the wire is sufficient. However, simple estimation reveals that the Joule heating in the resistive wire can become so large such that the wire could melt. To solve this issue, we propose to fabricate the Maxwell coil using REBCO (i.e., rare-earth barium copper oxide) superconducting tapes similar to those used in the work by Hahn \emph{et al.}~\cite{Hahn-2019-Nature}. A schematic of the proposed MLS setup is shown in Fig.~\ref{Fig5}~(a). A 24-T superconducting magnet with a bore diameter of 120 mm (existing at the NHMFL~\cite{Fu-2005-JMRI,Hendrickson-2015-JASMS}) is assumed for producing the $\mathbf{B}_0$ field. Four sets of gradient-field Maxwell coils made of REBCO pancake rings are placed in the bore of the superconducting magnet. Each pancake ring is made of 94 turns of the REBCO tape (width: 4 mm; thickness: 0.043 mm) so its cross section is nearly a square (i.e., 4 mm by 4 mm). The pancake rings are arranged along the diagonal lines of a standard gradient-field Maxwell coil and the averaged diameter of the pancake rings is about 8 cm. This coil configuration is found to produce a $\mathbf{B}_1$ field with minimal deviations from that of an ideal gradient-field Maxwell coil. While the superconducting magnet at the NHMFL is cooled by immersion in a liquid helium bath, the compact REBCO coils could be cooled conveniently by a 4-K pulse-tube cryocooler inside a shielded vacuum housing. A room-temperature center bore with a diameter as large as 6 cm can be used for sample loading and observation. When a current of about 290 A is applied in the REBCO tapes, a total turn-current $NI=4\times94\times290$~A$\simeq109$ kA can be achieved. Note that the quenching critical current of the REBCO tape can reach 700 A even under an external magnetic field of 30 T~\cite{Braccini-2010-SST}. Therefore, operating our REBCO coils with a tape current of 290 A should be safe and reliable.

\begin{figure*}[htb]
\centering
\includegraphics[width=0.95\linewidth]{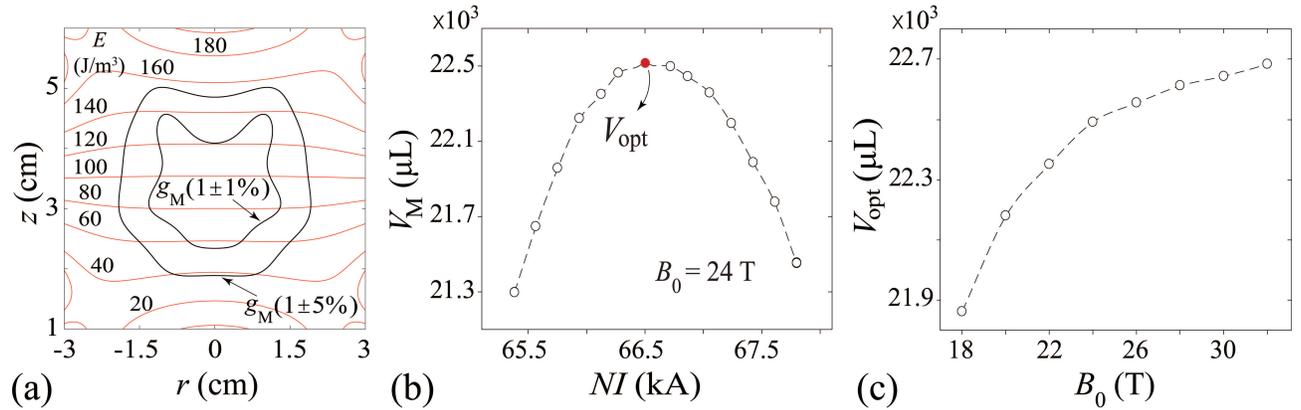}
\caption{(a) Contour plot of the specific potential energy $E(\mathbf{r})$ at $NI=66.55$ kA and $B_0=24$ T in the practical MLS. The black contours denote the boundaries of the regions in which the total force leads to an effective gravitational acceleration within 1\% and 5\% of $g_M$, respectively. (b) The functional volume $V_M$ in which the gravity varies within 5\% of $g_M$ versus the turn-current $NI$. (c) The peak volume $V_{\mathrm{opt}}$ versus $B_0$.}
\label{Fig6}
\end{figure*}

To prove the performance of the practical MLS design as depicted in Fig.~\ref{Fig5}~(a), we have repeated the previously presented optimization analyses. A representative plot of the specific potential energy $E(\mathbf{r})$ at a total turn-current $NI=108.37$ kA and $B_0=24$ T is shown in Fig.~\ref{Fig5}~(b). The overall shapes of the trapping region and the low-force region are nearly identical to those of the ideal gradient-field Maxwell coil. The dependance of $V_{1\%}$ on the turn-current $NI$ at $B_0=24$ T is shown in Fig.~\ref{Fig5}~(c). A peak functional volume $V_{\mathrm{opt}}$ about 3,450 $\mu L$ is achieved. In Fig.~\ref{Fig5}~(d), the peak volume $V_{\mathrm{opt}}$ obtained at various base field strength $B_0$ is shown. Again, the trend is similar to that in Fig.~\ref{Fig3}. Therefore, despite the change in the coil geometry as compared to the ideal gradient-field Maxwell coil, the performance of our practical design does not exhibit any significant degradation.

\subsection*{Emulating reduced gravities in extraterrestrial environment}\label{SecIIIb}
\noindent Beside levitating samples for near-zero gravity research, our MLS can also be tuned to partially cancel the Earth's gravity so that ground-based emulation of reduced gravities in extraterrestrial environment (such as on the Moon or the Mars) can be achieved. To demonstrate this potential, we present further analyses of the practical MLS shown in Fig.~\ref{Fig5} with lower applied currents for simulating the Martian gravity $g_M=0.38g$~\cite{Valles-2005-ASR}. In Fig.~\ref{Fig6}~(a), we show contour plots of the specific potential energy $E(\mathbf{r})$ for water samples in the practical MLS when a turn-current of $NI=66.55$ kA is applied at $B_0=24$ T. It is clear that the energy contour lines (red curves) are evenly spaced in the center region of the MLS, suggesting a fairly uniform and downward-pointing force in this region. We then calculate the magnitude of the force using Eq.~\ref{F-eqn}. The two black contours in Fig.~\ref{Fig6}~(a) represent the boundaries of the regions in which the total force leads to an effective gravitational acceleration within 1\% and 5\% of $g_M$, respectively. If we define the volume of the contour in which the gravity varies within 5\% of $g_M$ as our functional volume $V_\mathrm{M}$, its dependance on the turn-current at $B_0=24$ T is shown in Fig.~\ref{Fig6}~(b). This functional volume has a peak value $V_{\mathrm{opt}}$ of about $22.5\times10^3$ $\mu L$ at $NI=66.55$ kA. This peak volume is so large such that even small animals or plants can be accommodated inside. We have also calculated the peak volume $V_{\mathrm{opt}}$ at different base field strength $B_0$. As shown in Fig.~\ref{Fig6}~(c), initially the peak volume $V_{\mathrm{opt}}$ increases sharply with $B_0$, and then it gradually saturates when $B_0$ is greater than about 24 T. Operating the MLS at higher $B_0$ gives marginal gain in the functional volume.

\subsection*{Summary}\label{SecIIIc}
\noindent In conclusion, our analyses have clearly demonstrated the superiority of the proposed MSL concept in comparison with conventional solenoid MSLs. An unprecedentedly large and isotropic functional volume, i.e., about three orders of magnitude larger than that for a conventional solenoid MSL, can be achieved. The implementation of the superconducting magnet technology will also ensure stable operation of this MLS with a minimal energy consumption rate, which is ideal for future low-gravity research and applications.

\section*{Method}\label{SecIV}
\subsection*{Magnetic field calculation}\label{SecIV1}
\noindent The magnetic field $\mathbf{B}(\mathbf{r})$ generated at $\mathbf{r}$ by a current loop in three-dimensional space can be calculated using the Biot-Savart law~\cite{Jackson-1999-book}:
\begin{equation}
\mathbf{B}(\mathbf{r})=\frac{\mu_0I}{4\pi}\oint \frac{d\mathbf{l}\times(\mathbf{r}-\mathbf{l})}{|r|^3},
\end{equation}
where $d\mathbf{l}$ the elementary length vector along the current loop. For a field-gradient Maxwell coil with a radius $R=D/2$, the generated magnetic field $\mathbf{B}_1(\mathbf{r})$ can be decomposed into an axial component and a radial component due to the axial symmetry. If we set the z-axis along the co-axial line of the two loops and place the coordinate origin at the center of the bottom loop, the two components can be evaluated as:
\begin{eqnarray}
\resizebox{0.9\hsize}{!}{$
	\begin{split}
			&B_1^{(r)}(r,z)=\frac{\mu_0 I}{4\pi}\int_{0}^{2\pi}\left[\frac {Rz\cos(\phi)}{R_1^3}+\frac {R(L-z)\cos(\phi)}{R_2^3}\right]d\phi\\
			&B_1^{(z)}(r,z)=\frac{\mu_0 I}{4\pi}\int_{0}^{2\pi}\left[\frac {R^2 -Rr\cos(\phi)}{R_1^3}+\frac {Rr\cos(\phi)-R^2}{R_2^3}\right]d\phi\\
	\end{split}
$}
\end{eqnarray}
where
\begin{eqnarray}
	\begin{split}
		&R_1=\sqrt{[r-R\cos(\phi)]^2+ [R\sin(\phi)]^2+z^2}\\
		&R_2=\sqrt{[r-R\cos(\phi)]^2+ [R\sin(\phi)]^2+(z-L)^2}\\
	\end{split},
\end{eqnarray}
$L=\sqrt{3}D/2$ is the separation distance between the two loops, and $I$ is the current in each loop.

The magnetic field $\mathbf{B}_1(\mathbf{r})$ generated by the practical MLS design as depicted in Fig.~\ref{Fig5}~(a) can be calculated by superimposing the fields produced by the four sets of field-gradient Maxwell coils. The field of each coil is evaluated in the same way as outlined above. Counting in the base field $\mathbf{B}_0$, the total field is then given by $\mathbf{B}(\mathbf{r})=[B_0+B_1^{(r)}(\mathbf{r})]\mathbf{\hat{e}}_r+B_1^{(z)}(\mathbf{r})\mathbf{\hat{e}}_z$

For a solenoid with a length $L$ and a radius $R$, if we assume the wire is thin such that the turn number $N$ is large but the total turn-current $NI$ remains finite, an exact expression for the generated magnetic field can be derived based on the Biot-Savart law~\cite{Callaghan-1960-book,Caciagli-2018-JMMM}:
\begin{eqnarray}
\resizebox{0.9\hsize}{!}{$
\begin{split}
&B^{(r)}(r,z)=\frac{\mu_0 NI}{4\pi}\frac{2}{L}\sqrt{\frac{R}{r}}\left[\frac{k^2-2}{k}K(k^2)+\frac{2}{k}E(k^2)\right]^{\zeta_+}_{\zeta_-}\\
&B^{(z)}(r,z)=\frac{\mu_0 NI}{4\pi}\frac{1}{L\sqrt{Rr}}\left[\zeta k\left(K(k^2)+\frac{R-r}{R+r}\Pi(h^2,k^2)\right)\right]^{\zeta_+}_{\zeta_-}\\
\end{split}
$}
\end{eqnarray}
where
\begin{eqnarray}
	\begin{split}
		&k^2=\frac{4Rr}{(R+r)^2+\zeta^2}\\
		&h^2=\frac{4Rr}{(R+r)^2}\\
        &\zeta_\pm=z\pm L/2\\
	\end{split},
\end{eqnarray}
and the functions $K(k^2)$, $E(k^2)$, and $\Pi(h^2,k^2)$ are given by:
\begin{eqnarray}
	\begin{split}
		&K(k^2)=\int_0^{\pi/2}\frac{d\theta}{\sqrt{1-k^2\sin^2\theta}}\\
		&E(k^2)=\int_0^{\pi/2}d\theta\sqrt{1-k^2\sin^2\theta}\\
        &\Pi(h^2,k^2)=\int_0^{\pi/2}\frac{d\theta}{(1-h^2\sin^2\theta)\sqrt{1-k^2\sin^2\theta}}\\
	\end{split}.
\end{eqnarray}

\subsection*{Numerical method}\label{SecIV3}
\noindent The magnetic fields produced by the solenoid, the ideal gradient-field Maxwell coil, and the practical MLS design are all calculated using MATLAB. Considering the axial symmetry, we only evaluate the fields in the $r$-$z$ plane. The sizes of the computational domains for different types of designs are essentially shown in Fig.~\ref{Fig1}~(b), Fig.~\ref{Fig2}~(b), and Fig.~\ref{Fig5}~(b). Typically, the computational domain is discretized using a square grid with spatial resolutions $\Delta r=10$~$\mu$m and $\Delta z=10$~$\mu$m, which gives good convergence of the numerical results. The calculations assumed water properties at the ambient temperature, but the same procedures can be applied to other materials with different magnetic susceptibilities and densities.

\section*{Data Availability}
\noindent The computer codes and the data supporting the findings of this study are available from the corresponding author upon request.

\section*{Acknowledgments}
\noindent This work is supported by National Science Foundation under Grant No. CBET-1801780. The work was conducted at the National High Magnetic Field Laboratory, Florida State University, which is supported by National Science Foundation Cooperative Agreement No. DMR-1644779 and the state of Florida.

\section*{Author contributions}
\noindent W.G. designed the research; H.S. conducted the simulations; H.S. and W.G. analyzed the results and wrote the paper.

\section*{Competing interests}
\noindent The authors declare no competing interests

\section*{References}
\bibliographystyle{unsrt}
\bibliography{Levitation}

\end{document}